\documentclass{PoS}

\renewcommand\textfloatsep{5ex}

\title{Software correlators as testbeds for RFI algorithms}

\ShortTitle{Software correlators as testbeds}

\author{\speaker{Adam Deller}\\
        National Radio Astronomy Observatory\\
        E-mail: \email{adeller@nrao.edu}}

\abstract{In--correlator techniques offer the possibility of identifying and/or excising radio frequency interference (RFI) from interferometric observations at much higher time and/or frequency resolution than is generally possible with the final visibility dataset. Due to the considerable computational requirements of the correlation procedure, cross--correlators have most commonly been implemented using high--speed digital signal processing boards, which typically require long development times and are difficult to alter once complete.  ``Software" correlators, on the other hand, make use of commodity server machines and a correlation algorithm coded in a high--level language.  They are inherently much more flexible and can be developed -- and modified -- much more rapidly than purpose--built ``hardware" correlators.  Software correlators are thus a natural choice for testing new RFI detection and mitigation techniques for interferometers.  The ease with which software correlators can be adapted to test RFI detection algorithms is demonstrated by the addition of kurtosis detection and plotting to the widely used DiFX software correlator, which highlights previously unknown short--duration RFI at the Hancock VLBA station.}

\FullConference{RFI mitigation workshop - RFI2010,\\
		March 29-31, 2010\\
		Groningen, the Netherlands}
		
\begin{document}

\section{Interferometry, cross--correlators and RFI}
Radiotelescope arrays are inherently more robust against RFI than single dishes, due to the spatial separation between the array elements.  However, sources of RFI which are visible to more than one element will not be rejected during the correlation process and will corrupt the measured visibilities between those pairs of elements ({\em baselines}) and then the final interferometric image.  This problem is particularly severe for short baselines, where interfering sources are least likely to have been attenuated significantly before being received at neighbouring stations.

The cross--correlation algorithm used by radio interferometers to estimate the visibility function has been extensively reviewed elsewhere (see e.g. \cite{thompson94a} and \cite{romney99a}) and is not re--derived here.  The most important point of note is that cross--correlators can be built in an FX style (where the sampled baseband data is segmented and channelised on a per--station basis, and then cross-multiplied) or an XF style (in which the data streams for every baseline are cross--correlated with a range of lags, and the resultant lag spectrum is periodically Fourier transformed to generated the final visibility spectrum).  From the point of view of RFI detection and mitigation, the FX algorithm offers greater flexibility, since it provides access to the final channelised visibilities at higher time resolution.

Previous efforts to mitigate RFI in interferometric arrays have focused on identification and excision/correction of the RFI either before \cite{baan04a} or after \cite{athreya09a} the correlator.
Integrating RFI mitigation algorithms {\em inside} the correlator, however, exploits the advantages of both approaches:

\begin{enumerate}
\item Unlike post--correlation techniques, inspection of the data at high time resolution is possible, providing better signal--to--noise for the detection of faint, short--duration RFI; 
\item Likewise, if the duty cycle of the RFI is low, excision of the RFI at high time resolution is possible and will result in the loss of less good data;
\item Unlike pre-correlation techniques, no new steps are inserted into the serial signal processing chain, minimising the chance of introducing unwanted effects; and
\item Utilising the processing which is already undertaken in the correlator allows potentially more powerful techniques to be applied.
\end{enumerate}

However, incorporating RFI detection and mitigation inside the cross--correlator is more challenging than traditional approaches, since:

\begin{enumerate}
\item The data can only be inspected on--the--fly;
\item The implementation of the RFI algorithms cannot consume resources or otherwise impose constraints that would disturb the calculation of the visibility function; and
\item In most correlators, access to intermediate data products such as partially integrated visibility spectra is limited.
\end{enumerate}

Since software correlators can expose intermediate data products in a convenient (software--accessible) form, they are ideal for testing new RFI detection and mitigation techniques, with a much shorter development time than would be required on a hardware correlator.  The DiFX software correlator \cite{deller07a}, which is the subject of this investigation, is briefly described below.

\subsection{The DiFX correlator}

The Distributed FX (DiFX) software correlator is described in detail in \cite{deller07a}, and its relevant qualities for high time resoution RFI studies are briefly summarised here.  As the name suggests, it is an FX style correlator, which distributes data across multiple computing nodes using time--division multiplexing.  The length of the time slices (termed ``sub--integrations" below) is completely configurable, down to the inverse of the final spectral resolution, and the spectral resolution itself is completely configurable.  DiFX thus offers a very flexible platform for inspecting the RFI characteristics of interferometric data.  DiFX is used for production correlation of both the Very Long Baseline Array (VLBA) in the USA and the Long Baseline Array (LBA) in Australia, providing ample opportunities to test RFI algorithms on real data.

DiFX is written in C++, but uses optimised C vector libraries to carry out the bulk of correlation operations.  This allows both a convenient and comprehensible high--level design, and good performance.  Access to the vectors which hold (for example) the sampled baseband data and the visibility results is easily provided through pointers, and since the correlator runs in an unclocked fashion (in contrast to purpose--built hardware correlators using Field Programmable Gate Arrays or custom logic, which will generically be referred to as Digital Signal Processing -- DSP -- boards below) there are no timing constraints on the addition of extra RFI--related processing.

\section{Kurtosis--based RFI detection}
\label{sec:kurtosis}
Kurtosis measures the ``peakedness" of the probability distribution of a real--valued random variable.  It is defined as the fourth central moment of the variable divided by the square of the second central moment (where the second central moment is also the variance).  Closely related is the ``kurtosis excess", which is equal to the kurtosis minus 3.  The kurtosis of a normally distributed variable is equal to 3, and hence the kurtosis excess of a normally distributed variable is 0.

Since the explicit assumption of radio interferometry is that bona fide radio astronomical signals are normally distributed, the calculation of kurtosis provides a useful diagnostic of the presence of RFI.  A continuously present sinusoid has a kurtosis excess $<0$ (its distribution is {\em platykurtic}), while impulsive bursts lead to a kurtosis excess $>0$ (a {\em leptokurtic} distribution).  The effects of RFI with varying duty cycles on kurtosis in a radio astronomy context are investigated in \cite{deroo07a}.

In practice, the central moments of the signal are not known and must be estimated from a sample of data.  This is commonly achieved using unbiased estimators of the second and fourth cumulants (``k statistics") in place of the second and fourth central moments, respectively.  

In the ideal case of a perfect filterbank, each output spectral channel can be treated as an independent time series and standard time--domain kurtosis analysis can be applied.  However, this straightforward approach is complicated by the fact that cross-correlators window data in the time domain, produce complex channelised data, and result in non-negligible leakage between the spectral channels.  A more general concept of spectral kurtosis (SK; \cite{nita07a}) has been developed which accounts for these complications, while preserving the properties of kurtosis excess (Gaussian input data gives SK = 0, impulsive RFI yields SK > 0, and continuous sinusoids yield SK < 0).  The kurtosis calculation implemented in DiFX is somewhat simplified compared to the newly developed SK concept, but illustrates the principles of RFI detection and the ease with which the algorithm can be incorporated into a software correlator.

\section{Implementation of kurtosis detection in DiFX}
\enlargethispage{3ex}
As an FX style correlator, DiFX already produced channelised data for each antenna, which is available in the code before the cross-multiplication to form visibilities.  Over the course of a sub--integration, the 1st, 2nd, 3rd and 4th raw moments ($m_1$, $m_2$, $m_3$ and $m_4$) for each (complex) spectral channel were maintained.  At the end of each sub--integration, the 2nd and 4th central moments $u_2$ and $u_4$ were calculated from the raw moments, and the kurtosis for each channel calculated as $u_4/u_2^2$.  The calculated kurtosis values were written to a text file for later analysis.  The duration of the sub--integration can be (and was) varied to times less than a millisecond, using the existing software correlator infrastructure.

This simple approach yields a biased estimation of the kurtosis, for the reasons given in Section~\ref{sec:kurtosis} above.  At the time of implementation, the author was not aware of the SK concept, which is why this simple implementation was made.  Converting the existing simple implementation to the full SK calculation will be trivial, however, and is covered in Section~\ref{sec:conclusions}.  Nevertheless, Section~\ref{sec:results} shows that even a biased estimation of the kurtosis can be used to show the presence and spectral characteristics of RFI.

The greatest value of the kurtosis implementation in DiFX, however, is as a proof of concept for the rapid prototyping of RFI algorithms.  In total, $\sim$100 lines of code were necessary for the kurtosis calculation, bookkeeping and storage.  This is much less than would be required on a system based on DSP boards, due to the abstraction made possible by the use of high level languages.  Implementing and testing this additional code required only a few hours of coding.  Compared to the development which would be required on a DSP board correlator, this is extremely rapid.

\section{Results}
\label{sec:results}
To test the kurtosis implementation, a VLBA test experiment from 22 June 2009 at 1650 MHz was correlated using DiFX with kurtosis calculation enabled.  The spectrum around 1650 MHz is moderately contaminated with RFI at many VLBA sites.  The results below show a single 8 MHz band of right circular polarisation, spanning 1650.49 MHz to 1658.49 MHz.  Auto-- and cross--correlation spectra were integrated for a standard value of 2 seconds, while the kurtosis calculations were integrated for 100 ms, providing a higher time resolution view of the RFI.

\begin{figure}[b!]
\begin{center}
\begin{tabular}{cc}
\includegraphics[width=0.3\textwidth, angle=270]{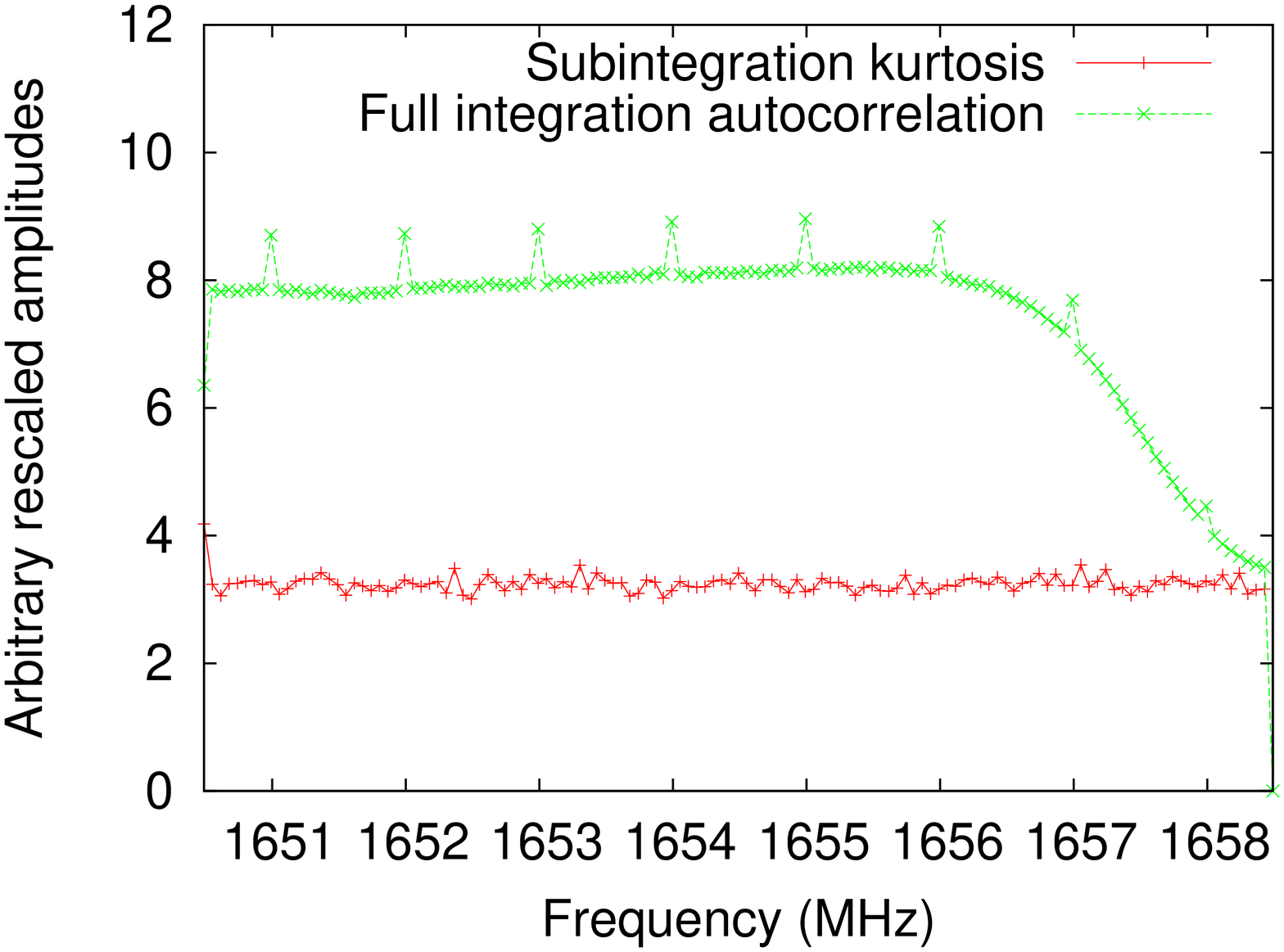} &
\includegraphics[width=0.3\textwidth, angle=270]{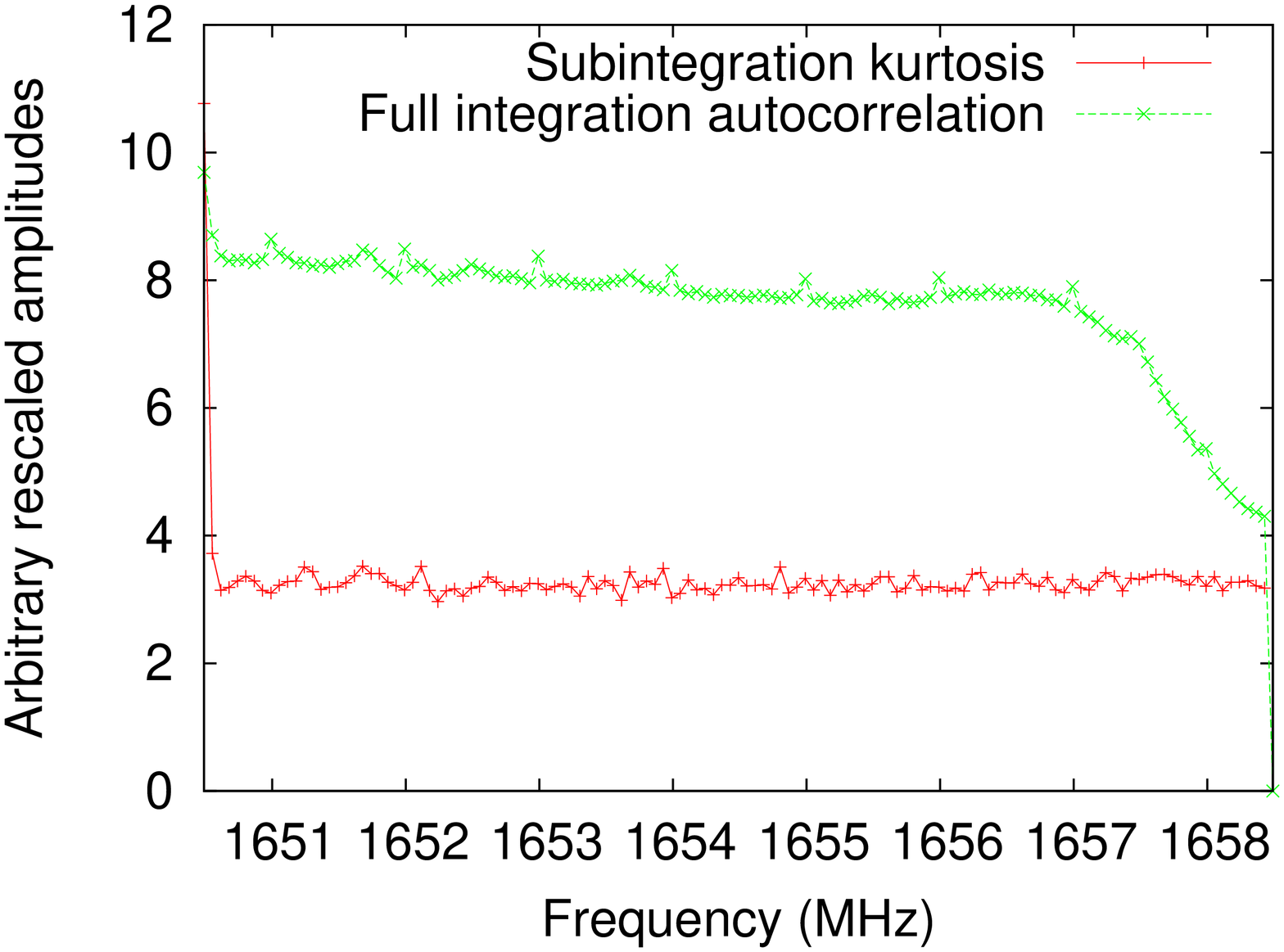} \\
\includegraphics[width=0.3\textwidth, angle=270]{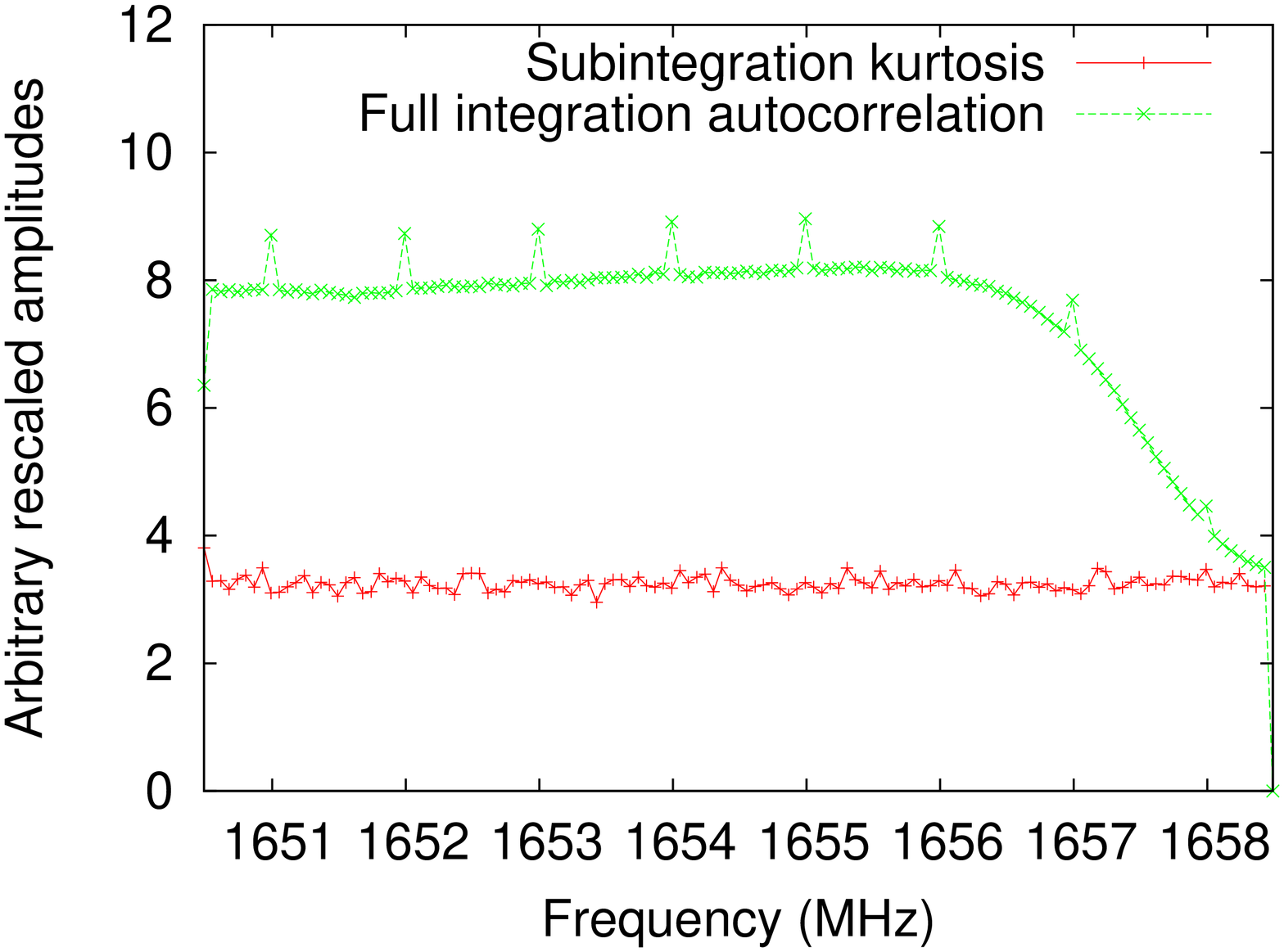} &
\includegraphics[width=0.3\textwidth, angle=270]{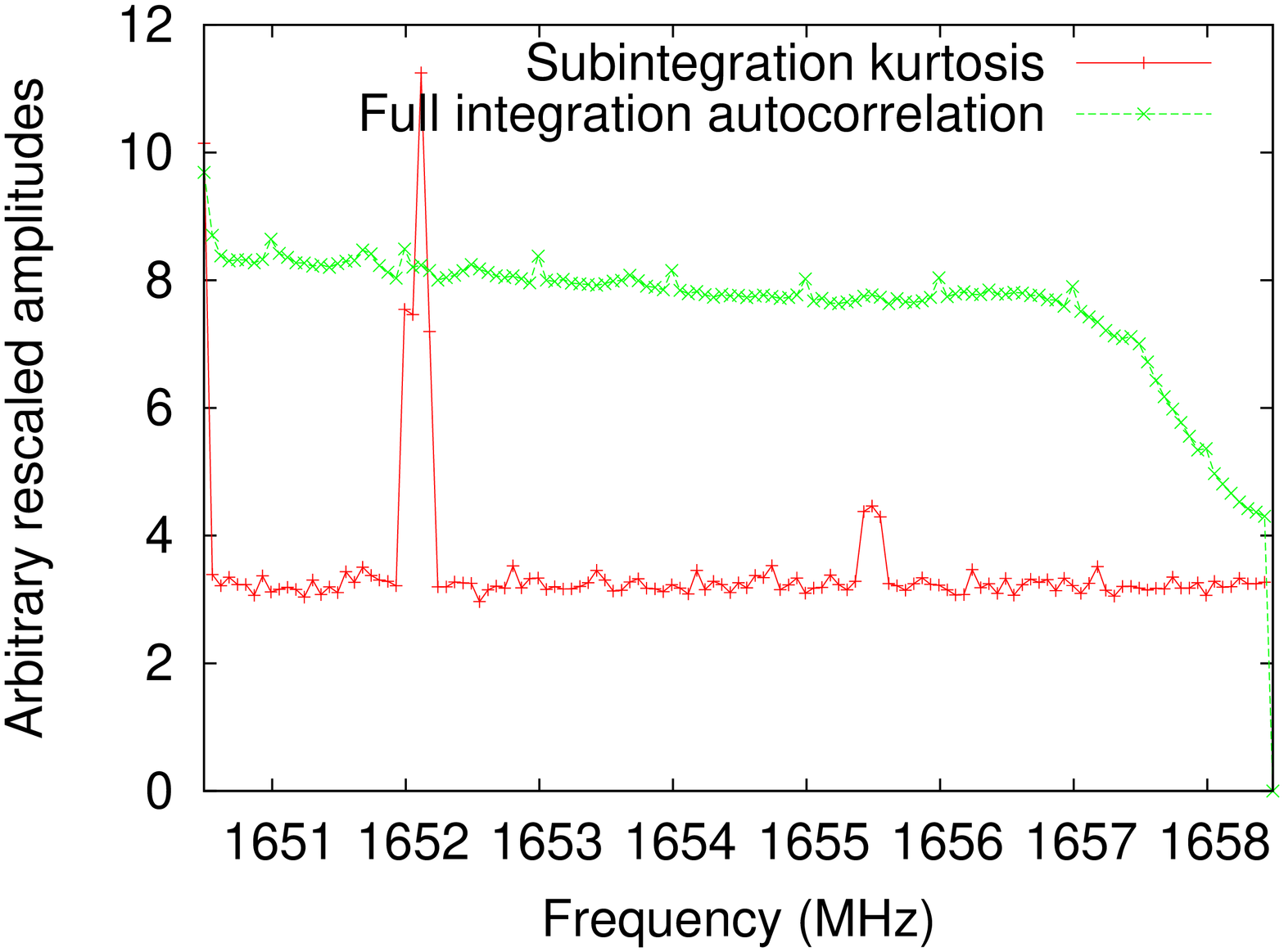} \\
\includegraphics[width=0.3\textwidth, angle=270]{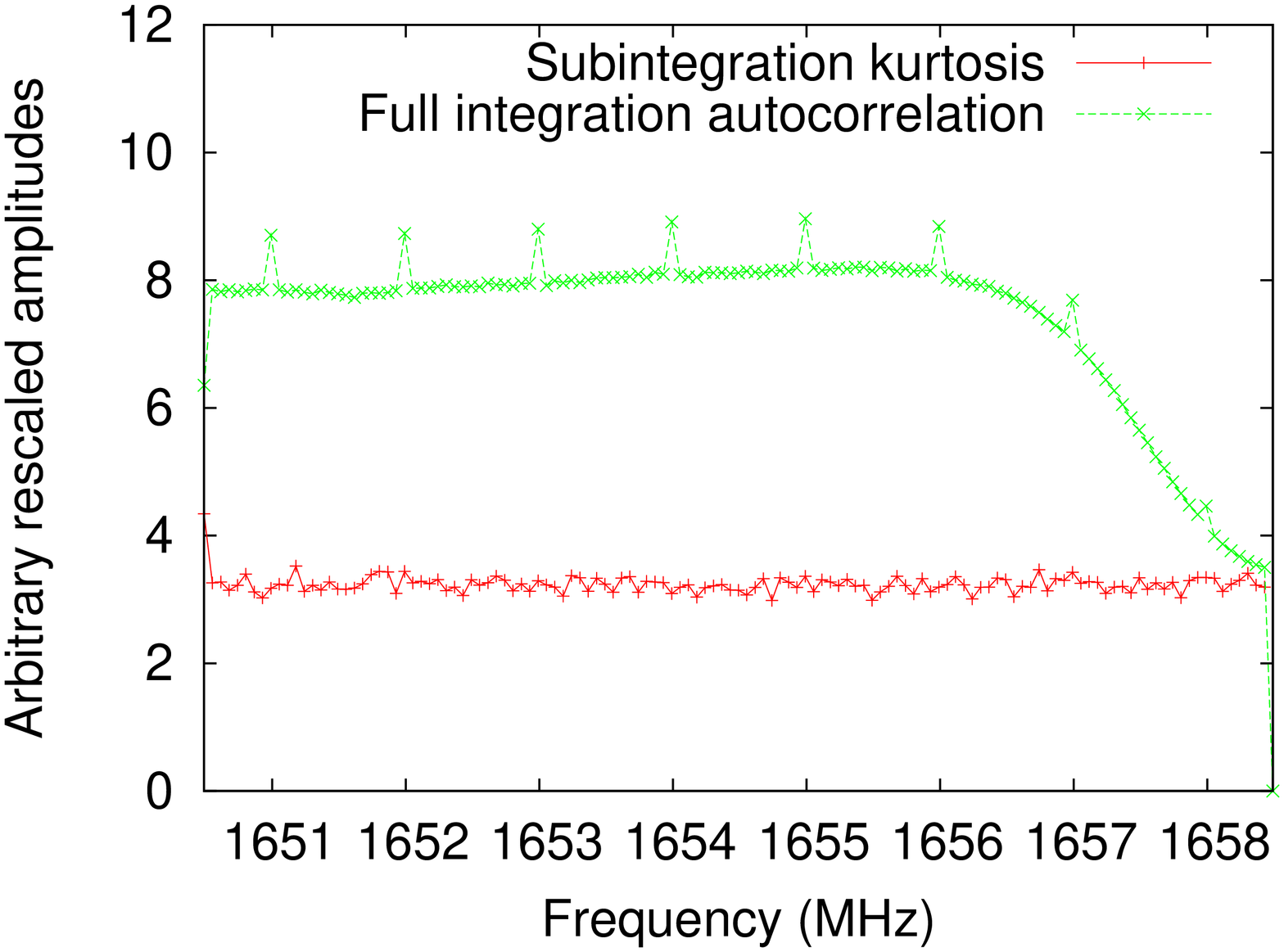} &
\includegraphics[width=0.3\textwidth, angle=270]{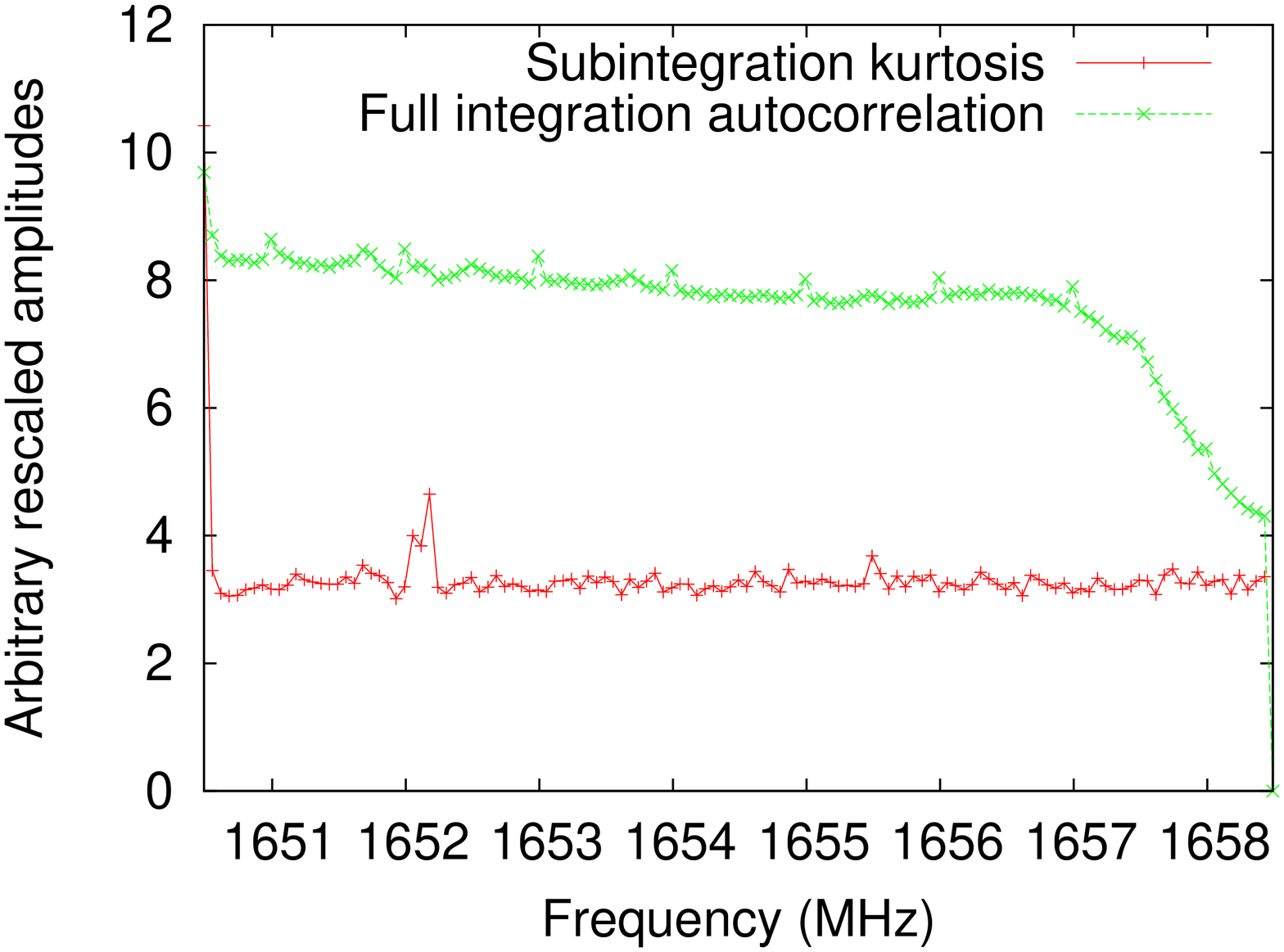} 
\end{tabular}
\caption{Autocorrelation (green) and kurtosis (red) for several successive 100 ms snapshots during the observation,
time running from top to bottom.  
The Fort Davis station is shown in the left hand column, with no discernible RFI.  The Hancock station is shown in 
the right hand column, and non-Gaussian signals are clearly present based on the kurtosis results.  The autocorrelation 
amplitudes in green have been arbitrarily scaled for convenient plotting on the same scale as the kurtosis measurements. 
The autocorrelation amplitudes remain constant, as they are taken from a 2 second integration covering this 300 ms period. 
Kurtosis (not kurtosis excess) is shown.  It is clear that the RFI varies dramatically on timescales shorter than 100 ms.
\label{fig:kurtosis}}
\end{center}
\end{figure}

Figure~\ref{fig:kurtosis} shows a sample of the results from two stations.  Three adjacent 100 ms kurtosis integrations are shown for the Fort Davis (left panel) and Hancock (right panel) stations, with the autocorrelations shown for comparison.  The autocorrelation spectra are relatively clean for both stations -- the spikes at 1 MHz intervals are the injected pulse calibration tones, used for removing instrumental delays.  Small deviations from the expected smooth bandpass are seen for the Hancock station, but the location and magnitude of the RFI is not obvious.  The kurtosis results, however, show several strong, rapidly varying RFI sources at Hancock.  Movies made from these plots, which show the varying RFI more clearly, are available at {\small \verb+http://www.aoc.nrao.edu/~adeller/rfi/+}.

The source of the RFI at the Hancock station is not known.  It is clearly variable on ms timescales and is somewhat spectrally confined, although it has a substantial bandwidth of around 0.5 MHz. Other Hancock bands exhibit similar characteristics, including in the protected 1660 -- 1670 MHz range.  When kurtosis--based blanking (see Section~\ref{sec:conclusions}) is implemented, the data which is clearly RFI--dominated could easily be excised, recovering most of the Hancock data.


The measured receiver noise temperature at Hancock has been observed to be erratic at the band encompassing these observations, and the knowledge gained from these tests has been passed on to NRAO staff to aid in the search for potential self--generated RFI, which might possibly originate in the noise injection circuit. Obtaining high signal to noise results on short timescales using the kurtosis output of DiFX, which would not possible with the normal cross and/or autocorrelations, has provided useful information for diagnosing this problem.

\enlargethispage{4ex}

\pagebreak
\section{Conclusions and future work}
\label{sec:conclusions}
A simple kurtosis estimator has been successfully implemented in the DiFX software correlator.  This initial foray into high time resolution in--correlator RFI detection was made with minimal effort, requiring only an afternoon of code development and testing, and provided useful information on previously unknown RFI at the Hancock VLBA station.  The present simplistic implementation should soon be updated to the unbiased SK estimator, and a thorough analysis of the effects of the coarse front--end quantisation should also be undertaken.  Ultimately, it may be possible to generalise the SK concept to the crosscorrelation outputs of the correlator.

With or without these improvements, however, the calculated kurtosis could be used to flag data on the fly (on timescales shorter than a full integration) or to produce a flag table which could be inspected and applied later in offline analysis.  The latter case, whilst allowing the flexibility to deselect some flags, loses the ability to save data which has been affected for only a short subset of a given integration.  In either case, this approach is still an excision procedure which ultimately results in data loss.  "Kurtosis blanking" and "kurtosis flagging" will be added as DiFX features in the near feature, and the usability and presentation of the saved kurtosis results will be improved.

Development of other forms of in--correlator RFI mitigation are planned for the DiFX correlator.  One example is the rejection of bright sources outside the desired array field of view, which can be achieved through field of view shaping using tapered time integrations (see \cite{lonsdale04a}).  This will allow a determination of the relative trade-off between accuracy of the tapering function and the performance overhead incurred.  The success of the kurtosis development detailed in this work indicates that these approaches can be rapidly and effectively implemented.

\end{document}